\documentclass[aps,preprint,prd,epsfig,nofootinbib]{revtex4-1}
\pdfoutput=1
\usepackage{graphicx}
\usepackage{longtable}
\usepackage{float}
\usepackage{dcolumn}
\usepackage{bm}
\usepackage{appendix}
\usepackage{multirow}
\usepackage{color}
\usepackage[utf8]{inputenc}
\usepackage{footmisc}
\usepackage{soul}
\usepackage{txfonts}
\usepackage{xcolor}
\usepackage{graphicx}
\usepackage{bm}
\usepackage{ulem}

\usepackage{amsmath}

\usepackage[colorlinks=true,linkcolor=blue,citecolor=blue]{hyperref}
\usepackage{feynmp-auto}
\usepackage{fancyhdr}
\usepackage{lineno}

\usepackage{wasysym}

\usepackage{xcolor}
\usepackage{graphicx}
\usepackage{dcolumn}
\usepackage{bm}
\usepackage{amssymb}
\usepackage{blindtext}
\usepackage{hyperref}
\begin{document}
\title { Scalar contributions to the S, T, U parameters in a 3-3-1 model}
\author{A. Doff$^{\ddag}$\footnote[1]{\href{mailto:agomes@utfpr.edu.br }{\footnotesize agomes@utfpr.edu.br }}}
\author{C. A. de S. Pires$^{\dagger}$\footnote[2]{\href{mailto:cpires@fisica.ufpb.br}{\footnotesize cpires@fisica.ufpb.br}}}

\affiliation{\footnotesize $^{\ddag}$ Universidade Tecnologica Federal do Parana - UTFPR - DAFIS, R. Doutor Washington Subtil Chueire, 330 - Jardim Carvalho, 84017-220, Ponta Grossa, PR, Brazil}
\affiliation{\footnotesize $^{\dagger} $Departamento de F\'isica, Universidade Federal da Para\'iba, Caixa Postal 5008, 58051-970, Jo\~ao Pessoa, PB, Brazil}
\date{\today}
\begin{abstract}
Electroweak precision tests, expressed through the oblique parameters $S$, $T$, and $U$, impose stringent constraints on physics beyond the Standard Model. Gauge extensions of the Standard Model based on the $SU(3)_L \times U(1)_N$ symmetry  predict a rich scalar and gauge spectrum that contribute to these parameters. Previous studies have shown that 3-3-1 gauge bosons   give negligible contributions to  the oblique parameters, while the contributions of the scalar sector to these parameters have received comparatively little attention. In particular, for the version of the $SU(3)_L \times U(1)_N$
 model with right-handed neutrinos, the impact of the scalar sector on $S$,  $T$
, and  $U$ has not yet been  addressed. In this work, we fill this gap  and address sistematically the scalar contributions to the $S$, $T$ and $U$ within this version. As main result, we show that the parameter $T$ put stringent constraints on the masses and energy scales associated to the spectrum of scalars  of the model.
\end{abstract}
\maketitle
\section{Introduction}
Electroweak precision tests of the Standard Model (SM), encapsulated in the oblique parameters $S$, $T$, and $U$, provide some of the most stringent probes of new physics. Defined through vacuum polarization corrections to the electroweak gauge bosons, these parameters are highly sensitive to custodial symmetry breaking by means of extensions of the SM that introduce new particle content, particularly in the form of electroweak doublets.

Gauge extensions of the Standard Model (SM) based on the symmetry group ${\cal G}_{331}=SU(3)_C \times SU(3)_L \times U(1)_N$ (3-3-1)\cite{Singer:1980sw, Frampton:1992wt,Pisano:1992bxx,Foot:1994ym}
 constitute an appealing framework for physics beyond the SM. These models address several longstanding issues in particle physics, including the problem of family replication\cite{Liu:1993gy,Pisano:1996ht}, the quantization of electric charge\cite{deSousaPires:1998jc,deSousaPires:1999ca}, and the strong CP problem\cite{Pal:1994ba}. From a phenomenological perspective, a distinctive feature of such models is the inevitable emergence of flavor-changing neutral currents at tree level, mediated by new scalars and gauge bosons\cite{Long:1999ij,Buras:2014yna, Oliveira:2022vjo,Escalona:2025rxu, Escalona:2025jla}. Regarding their impact on electroweak precision observables, it is noteworthy that in their minimal versions as scalars as gauge bosons contribute nontrivially to the oblique parameters. However, few works have addressed this issue within the various versions of the 3-3-1 models\cite{Sasaki:1992np,Liu:1993fwa,Frampton:1997in,Long:1999bny,Rehman:2025urc}.  In light of this, we extend the analysis presented in Ref. \cite{Long:1999bny}, which focused exclusively on the gauge boson contributions to the oblique parameters, by calculating the contributions of the scalar sector of the 3-3-1 model with right-handedn neutrinos(331RHN) to the $S$, $T$ $U$ parameters. 
 
 To determine whether  an extension of the SM contribute to the $S$, $T$, $U$ parameters, one must identify the residual  ${\cal G}_{SM}=SU(3)_C \times SU(2)_L \times U(1)_Y$ gauge content after symmetry breaking. In the specific  case of the 3-3-1 models, the spontaneous breaking of ${\cal G}_{331}  \to {\cal G}_{SM}$ leaves behind a  mult-Higgs sector  together with  a gauge doublet formed by the  bileptons gauge bosons. In the original version of the 331RHN, the symmetry breaking leaves behind three doublets of scalars(3HDM), one of which is inert. We therefore calculate the contributions of this 3HDM to the oblique parameters and extract relations among the scalar masses and energy parameters intrinsic to the 331RHN. The parameters in questions are the energy scale associated to the ${\cal G}_{331}$ symmetry breaks ($v_{\chi^{\prime}}$) and the energy parameter associated to the trilinear coupling ($f$) that composes the scalar potential. Our analysis shows that the oblique parameter $T$
 enforces tight correlations between $v_{\chi^{\prime}}$ and 
$f$, yielding the first significant constraint on  $f$  in the literature.

This work is organized as follows. In Sec. II we introduce key aspects of the model that provide the foundation for the problem under study. In Sec. III, we compute the scalar contributions to the oblique parameters. In  Sec. IV we present our numerical results, and in Sec. V we  conclude with a summary of our findings and final remarks.

\section{Some aspects of the model}
\subsection{Fermion content}

The lepton content comprises triplets and singlets under $SU(3)_C \times SU(3)_L \times U(1)_N$,
\begin{eqnarray}
L_{a_L} = \left (
\begin{array}{c}
\nu_a  \\
e_a\\
\nu^c_a
\end{array}
\right)_L \sim (1,3,-1/3) \,,\,\,\, e_{a_R}\sim(1,1,-1), \,
\label{lc}
 \end{eqnarray}
with $a=e\,,\,\mu\,,\,\tau$ representing the three SM generations of leptons.

The quark sector  comes in three families. Here anomaly cancellation requires that at least one family of quarks transform differently from the other two. So we have three variants with each one leading to different physical results\cite{Oliveira:2022vjo}. Each variant of the model leads to different physics effects with the current status of new physics hitting  that the third family transforming as triplet as the prefered case. Then we restrict our analisis here for this case which means the families of quarks  transforming by the 3-3-1 symmetry in the following way
\begin{eqnarray}
&&Q_{i_L} = \left (
\begin{array}{c}
d_{i} \\
-u_{i} \\
d^{\prime}_{i}
\end{array}
\right )_L\sim(3\,,\,\bar{3}\,,\,0)\,,u_{iR}\,\sim(3,1,2/3),\,\,\,\nonumber \\
&&\,\,d_{iR}\,\sim(3,1,-1/3)\,,\,\,\,\, d^{\prime}_{iR}\,\sim(3,1,-1/3),\nonumber \\
&&Q_{3L} = \left (
\begin{array}{c}
u_{3} \\
d_{3} \\
u^{\prime}_{3}
\end{array}
\right )_L\sim(3\,,\,3\,,\,1/3),u_{3R}\,\sim(3,1,2/3),\nonumber \\
&&\,\,d_{3R}\,\sim(3,1,-1/3)\,,\,u^{\prime}_{3R}\,\sim(3,1,2/3), \, \label{quarks}
\end{eqnarray}
where  the index $i=1,2$ is restricted to only two generations. The negative signal in the anti-triplet $Q_{i_L}$ is just to standardise the signals of the charged current interactions with the gauge bosons.  The primed quarks are new heavy quarks with the usual $(+\frac{2}{3}, -\frac{1}{3})$ electric charges. 

Observe that, in view of the spontaneous breaking ${\cal G}_{331}  \to {\cal G}_{SM}$, the triplets of fermions decompose in the standard doublets plus new fermions in the singlet form. Due to this the fermionic content does not provide any new contributions to the oblique parameters.
\subsection{Scalar sector}

 The scalar sector involves three  scalar triplets,
\begin{eqnarray}
\eta = \left (
\begin{array}{c}
\eta^0 \\
\eta^- \\
\eta^{\prime 0}
\end{array}
\right ),\,\rho = \left (
\begin{array}{c}
\rho^+ \\
\rho^0 \\
\rho^{\prime +}
\end{array}
\right ),\,
\chi = \left (
\begin{array}{c}
\chi^0 \\
\chi^{-} \\
\chi^{\prime 0}
\end{array}
\right ),
\label{tripletscalar} 
\end{eqnarray}
with $\eta$ and $\chi$ transforming as $(1\,,\,3\,,\,-1/3)$
and $\rho$ as $(1\,,\,3\,,\,2/3)$.

The most economical scalar potential formed by this set of scalar triplets involve the following terms
\begin{eqnarray} 
V(\eta,\rho,\chi)&=&\mu_\chi^2 \chi^2 +\mu_\eta^2\eta^2
+\mu_\rho^2\rho^2+\lambda_1\chi^4 +\lambda_2\eta^4
+\lambda_3\rho^4+ \nonumber \\
&&\lambda_4(\chi^{\dagger}\chi)(\eta^{\dagger}\eta)
+\lambda_5(\chi^{\dagger}\chi)(\rho^{\dagger}\rho)+\lambda_6
(\eta^{\dagger}\eta)(\rho^{\dagger}\rho)+ \nonumber \\
&&\lambda_7(\chi^{\dagger}\eta)(\eta^{\dagger}\chi)
+\lambda_8(\chi^{\dagger}\rho)(\rho^{\dagger}\chi)+\lambda_9
(\eta^{\dagger}\rho)(\rho^{\dagger}\eta) \nonumber \\
&&-\frac{f}{\sqrt{2}}\epsilon^{ijk}\eta_i \rho_j \chi_k +\mbox{H.c.}\,.
\label{potentialII}
\end{eqnarray}
This potential preserves lepton number and is invariant by the discrete $Z_2$ symmetry with $\eta\,,\,\rho \to -(\eta\,,\,\rho)$.

By simplification reasons, we restrict our approach to the case in which  only $\eta^0 \,\,, \rho^0$ and $\chi^{\prime 0}$ acquire vacuum expectation values (VEVs). We then shift these neutral fields  in the usual way
\begin{equation}
    \eta^0 =\frac{1}{\sqrt{2}}(v_{\eta}+R_{\eta}+iI_{\eta})\,\,,\,\,
    \rho^0 =\frac{1}{\sqrt{2}}(v_{\rho}+R_{\rho}+iI_{\rho})\,\,,\,\,
    \chi^{\prime 0} =\frac{1}{\sqrt{2}}(v_{\chi^{\prime}}+R_{\chi^{\prime }}+iI_{\chi^{\prime }})\,.
\end{equation}

The potential above provides the following set of  minimal constraint equations 
\begin{eqnarray}
 &&-\frac{f v_{\chi} v_{\rho}}{2 v_{\eta}}+ \lambda_2 v_{\eta}^2+\frac{1}{2} \lambda_4 v_{\chi}^2  +\frac{1}{2} \lambda_6  v_{\rho}^2+ \mu_\eta^2 =0 \,, \\
&&-\frac{f v_{\chi} v_{\eta}}{2 v_{\rho}}+\lambda_3 v_{\rho}^2+\frac{1}{2} \lambda_5 v_{\chi}^2 +\frac{1}{2} \lambda_6 v_{\eta}^2 + \mu_\rho^2 =0 \,,
\nonumber \\
&&
-\frac{f v_{\eta} v_{\rho}}{2v_{\chi}}+ \lambda_1 v_{\chi^0}^2 +\frac{1}{2} \lambda_4  v_{\eta}^2+\frac{1}{2} \lambda_5  v_{\rho}^2+  \mu_\chi^2  =0 \,.
\label{mincond} 
\end{eqnarray}

The mass matrix for the CP-even neutral scalars in the basis $(R_{\chi'}, R_\eta, R_\rho)$ is
\begin{equation}
M_R^2=
\begin{pmatrix}
\lambda_1 v^2_{\chi^{\prime}}+fv_\eta v_\rho/4v_{\chi^{\prime}} & \lambda_4 v_{\chi^{\prime}}  v_\eta/2- f v_\rho/4 &  \lambda_5 v_{\chi^{\prime}}  v_\rho/2- f v_\eta/4\\
\lambda_4 v_{\chi^{\prime}}  v_\eta/2- f v_\rho/4 & \lambda_2 v^2_\eta+ fv_{\chi^{\prime}} v_\rho/4v_\eta & \lambda_6 v_\eta v_\rho/2-f v_{\chi^{\prime}}/4\\
\lambda_5 v_{\chi^{\prime}}  v_\rho/2- f v_\eta/4 &  \lambda_6 v_\eta v_\rho/2-f v_{\chi^{\prime}}/4 &  \lambda_3 v^2_\rho+ fv_{\chi^{\prime}} v_\eta/4v_\rho
\end{pmatrix}.
\label{masseven}
\end{equation}
In what follows, we assume $R_{\chi^{\prime}}$ decouples from $R_\eta$ and $R_\rho$\footnote{This decoupling limit is physically motivated when $v_{\chi'} \gg v_\eta, v_\rho$,  ensuring that heavy 3-3-1 physics decouples from the electroweak scale.}. This requires
\begin{equation}
     \lambda_4 v_{\chi^{\prime}}  v_\eta/2- f v_\rho/4=0\, \quad \text{ and }\, \quad \lambda_5 v_{\chi^{\prime}}  v_\rho/2- f v_\eta/4=0.
\end{equation}
In this limit the $3\times 3$ mass matrix above reduces to the $2\times 2$ mass matrix that in the basis $( R_\eta,R_\rho)$ takes the
\begin{equation}
M_H^2=
\begin{pmatrix}
 \lambda_2 v^2_\eta+ fv_{\chi^{\prime}} v_\rho/4v_\eta & \lambda_6 v_\eta v_\rho/2-f v_{\chi^{\prime}}/4\\
  \lambda_6 v_\eta v_\rho/2-f v_{\chi^{\prime}}/4 &  \lambda_3 v^2_\rho+ fv_{\chi^{\prime}} v_\eta/4v_\rho
\end{pmatrix}
\label{masseven2}
\end{equation}
 
After diagonalization, the CP-even eigenstates are
\begin{eqnarray}
    &&h=\cos \varphi R_\eta +  \sin \varphi R_\rho \nonumber \\
    && H= \cos \varphi R_\rho -  \sin \varphi R_\eta \nonumber \\
   && H^{\prime}= R_{\chi^{\prime}}.
\end{eqnarray}
Here, $H'$ is a  CP-even scalar with mass given by $m^2_{H^{\prime}}= \lambda_1v^2_{\chi'}$. The remaining CP-even scalars are $h$ (the SM-like Higgs) and $H$, whose mass depends on $f$ and $v_{\chi'}$. The angle $\varphi$ diagonalizes $M_H^2$ via a $2\times2$ rotation. For the explicit diagonalization procedure, we refer the reader to Refs. \cite{Long:1997vbr,Tully:1998wa,Ponce:2002sg,Diaz:2003dk,Palcu:2013sfa,Pinheiro:2022bcs}.

Next, we consider the CP-odd scalars. Considering the basis $(I_{\chi^{\prime}}, I_\eta, I_\rho)$, the mass matrix is 
\begin{equation}
M_I^2=\frac{f}{4}
\begin{pmatrix}
v_\eta v_\rho/v_{\chi^{\prime}} &   v_\rho &   v_\eta\\
 v_\rho &  v_{\chi^{\prime}} v_\rho/ v_\eta &  v_{\chi^{\prime}}\\
  v_\eta &   v_{\chi^{\prime}} &  v_{\chi^{\prime}} v_\eta/v_\rho
\end{pmatrix}.
\label{MI}
\end{equation}
This matrix can be diagonalized analytically.  Assuming $v_{\chi^{\prime}}\gg v_\eta\,,\,v_\rho$, we obtain two zero eigenvalues, which correspond to the eigenstates $I_{\chi^{\prime}}$ and $G$, and a massive eigenstate $A$ given by  
\begin{eqnarray}
 && G= \cos \phi I_\eta -\sin \phi I_\rho \nonumber \\
  &&A= \cos \phi I_\rho  +\sin \phi I_\eta ,
  \label{DefAbas}
\end{eqnarray}
where $\tan \phi = \frac{v_\rho}{v_\eta}$.

We now discuss the charged scalars. In the basis $(\chi^-, \rho^{\prime -}, \eta^- , \rho ^-)$, the mass matrix is given by:
\begin{equation}
M_C^2=\frac{1}{2}
\begin{pmatrix}
\lambda_8 v^2_\rho+f v_\eta v_\rho / v_{\chi^{\prime}} & \lambda_8 v_\rho v_{\chi^{\prime}} +fv_\eta & 0 & 0 \\
\lambda_8 v_\rho v_{\chi^{\prime}} +fv_\eta & \lambda_8 v^2_{\chi^{\prime}} +f v_\eta v_{\chi^{\prime}}/ v_\rho & 0 & 0\\
0 & 0 & \lambda_9 v^2_\rho  +f v_\rho v_{\chi^{\prime}}/ v_\eta & \lambda_9 v_\rho v_\eta  +f  v_{\chi^{\prime}} \\
0 & 0 &  \lambda_9 v_\rho v_\eta  +f  v_{\chi^{\prime}} & \lambda_9 v^2_\eta  +f v_\eta v_{\chi^{\prime}}/ v_\rho
\end{pmatrix}.
\label{M+}
\end{equation}
We note that $\chi^-$ and $\rho^{\prime -}$ also carry two units of lepton number each. Thus, conservation of the lepton number prevents that $\chi^-$ and $\rho^{\prime -}$ mix with $\eta^- , \rho ^-$. After diagonalizing this matrix, we obtain two Goldstone bosons: $G_1^+$ and $ G_2^{ +}$, which are absorbed by $W^{\pm}$ and $W^{\prime \pm}$. The remaining two heavy charged scalars are $h_1^+ $  and $h_2^+ $ .The eigenstates are given by
\begin{eqnarray}
  && G_1^+=\cos \phi  \eta^+ - \sin \phi \rho^+ ,   \nonumber \\
  && h^+_1 = \cos \phi \rho ^+ + \sin \phi \eta^+ , \nonumber \\
   && G_2^{ +}= \cos \zeta\chi^+ +\sin \zeta\rho^{\prime +}, \nonumber \\
   && h_2^+=-\sin \zeta \chi^+ +\cos \zeta\rho^{\prime +},
\end{eqnarray}
where $\cos \zeta =\frac{v_{\chi^{\prime}}}{\sqrt{v^2_\eta + v^2_{\chi^{\prime}}}}$ and $\sin \zeta =\frac{v_{\eta}}{\sqrt{v^2_\eta + v^2_{\chi^{\prime}}}} $. 

The 331RHN has two other neutral scalars, $\chi^0$ and $\eta^{\prime 0}$. Both carry two units of lepton number. Assuming lepton number conservation, they do not mix with $\eta^0$, $\rho^0$, or $\chi^{\prime 0}$. In the  basis $(\chi^{ 0} , \eta^{\prime 0})$ the mass matrix reads:
\begin{equation}
M^2_{\chi \eta^{\prime}}=\frac{1}{4}
\begin{pmatrix}
\lambda_7v^2_\eta +f v_\eta v_\rho/v_{\chi^{\prime}} & -\lambda_7 v_\eta v_{\chi^{\prime}}-fv_\rho\\
-\lambda_7 v_\eta v_{\chi^{\prime}}-fv_\rho &  \lambda_7 v^2_{\chi^{\prime}} +f v_\rho v_{\chi^{\prime}}/v_\eta
\end{pmatrix}.
\label{MI2}
\end{equation}

After diagonalizing this mass matrix, we obtain  $G_3=\cos \zeta\chi^0 +\sin \zeta\eta^{\prime 0}$, which is a Goldstone boson absorbed by the non-Hermitian gauge bosons $U^0$ and $U^{0 \dagger}$. The other neutral scalar is  $H^{\prime \prime}=-\sin \zeta\chi^0 +\cos \zeta\eta^{\prime 0}$.

In summary, the scalar spectrum of the 331RHN may be divided in two block, namely $h$, $H$, $A$, $h^{\pm}_1$, and a second block  composed by $H^\prime$, $H^{\prime \prime}$, $h^{\pm}_2$. The mass spectrum of these scalars will be discussed below.

\subsection{Gauge Sector}
The gauge sector of the 331RHN comprises eight gluons associated to the gauge group $SU(3)_C$, eight gauge bosons associated to SU$(3)_L$, namely $W^\mu_1,\dots,W^\mu_8$, and a single gauge boson, $W^\mu_N$, associated to U$(1)_N$. After the spontaneous breaking of the symmetry the gauge bosons of the $SU(3)_L\times U(1)_N $ symmetry  mix among themselves to produces the four standard electroweak bosons along with five additional heavy states\cite{Long:1995ctv}. To obtain the corresponding gauge boson spectrum, we expand the scalar kinetic term,
\begin{equation}
\sum_{\Phi=\eta,\rho,\chi}({\cal D}^\mu \Phi)^{\dagger} ({\cal D}_\mu \Phi),
\end{equation}
where the covariant derivative is defined as
\begin{equation}
    {\cal D}_\mu = I\partial_\mu  -igW^a_\mu T^a-ig_N N W^N_\mu I,
\end{equation}
where $T^a= \frac{\lambda^a}{2}$ ($\lambda^a$ being the Gell-mann matrices) with $a=1,2,...,8$, $g$ is the SU$(3)_L$ gauge coupling and $g_N$ is the coupling associated to U$(1)_N$. The matrix $W^a_\mu T^a$ is given by
\begin{equation}
W^a_\mu T^a=\frac{1}{2}
\begin{pmatrix}
W^3_\mu + \frac{1}{\sqrt{3}}W
^8_\mu & W^1_\mu -iW^2_\mu &  W^4_\mu -iW^5_\mu \\
W^1_\mu +iW^2_\mu &- W^3_\mu + \frac{1}{\sqrt{3}}W^8_\mu & W^6_\mu -iW^7_\mu\\
 W^4_\mu +iW^5_\mu & W^6_\mu +iW^7_\mu &  -\frac{2}{\sqrt{3}}W^8_\mu
\end{pmatrix}.
\label{mass_gauge}
\end{equation}
From this kinetic term, after the spontaneous symmetry breaking of ${\cal G}_{331}  \to {\cal G}_{SM}$, we obtain the following set of eigenstates:  
\begin{eqnarray}
    &&W^1_\mu\,\,\,,\,\,\, W^2_\mu\,\,\,,\,\,\, W^{\prime \pm}_\mu=\frac{W^6_\mu \mp iW^7_\mu}{\sqrt{2}}\,\,\,,\,\,\,U^0_\mu=\frac{W^4_\mu-iW^5_\mu}{\sqrt{2}},\nonumber \\
    &&U^{0 \dagger}_\mu=\frac{W^4_\mu+iW^5_\mu}{\sqrt{2}}\,\,\,,\,\,\, W^3_\mu\,\,\,,\,\,\, B_\mu\,\,\,,\,\,\, Z^{\prime}_\mu,
\end{eqnarray}
where 
\begin{equation}
     B_\mu=-\frac{t_W}{\sqrt{3}}W^8_\mu +\sqrt{1-\frac{t^2_W}{3}}W^N_\mu \quad \quad \text{and} \quad \quad Z^{\prime}_\mu=\sqrt{1-\frac{t^2_W}{3}}W^8_\mu + \frac{t_W}{\sqrt{3}}W^N_\mu\,,
\end{equation}
and $t_W = \tan \theta_W$, with $\theta_W$ being the Weinberg angle. The neutral gauge boson $B_\mu$ is associated with the hypercharge generator of the U$(1)_Y$ abelian group.
Then, after the electroweak symmetry breaking of ${\cal G}_{SM}  \to \text{U}(1)_{\text{em}}$, the $W^1_\mu$ boson mixes with $W^2_\mu$ to form the  standard physical charged gauge bosons, $W^{\pm}_\mu=\frac{W^1_\mu \mp iW^2_\mu}{\sqrt{2}}$, and $W^3_\mu$ combines with $B_\mu$ to compose the photon and the standard massive neutral gauge boson, $A_\mu =s_W W^3_\mu +c_W B_\mu$ and $Z_\mu=c_W W^3_\mu -s_W B_\mu$, where $s_W = \sin \theta_W$ and $c_W = \cos \theta_W$. 

In summary, the 331RHN model contains the following spectrum of gauge bosons: 
\begin{equation}
W^{\pm}_\mu\,\,\,,\,\,\, W^{\prime \pm }_\mu\,\,\,,\,\,\,U^0_\mu \,\,\,,\,\,\, U^{0 \dagger}_\mu \,\,\,,\,\,\, A_\mu \,\,\,,\,\,\,Z_\mu\,\,\,,\,\,\,Z^{\prime}_\mu,
\end{equation}
whose masses correspond to
\begin{eqnarray}
    && m^2_W=\frac{g^2}{4}(v^2_\eta + v^2_\rho)\,\,\,,\,\,\, m^2_{W^{\prime}}=\frac{g^2}{4}(v^2_\eta + v^2_{\chi^{\prime}})\,\,\,,\,\,\,m^2_{U^0}=\frac{g^2}{4}(v^2_\rho + v^2_{\chi^{\prime}}),\nonumber \\
    &&  m^2_Z=\frac{g^2}{4c^2_W}(v^2_\eta + v^2_\rho)\,\,\,,\,\,\,  m^2_{Z^{\prime}}=\frac{g^2}{3-4s^2_W}  v^2_{\chi^{\prime}} \,\,\,,\,\,\, m_A^2 =0\,\,\,.
    \label{GBmass}
\end{eqnarray}
From these expression, we infer that
\begin{equation}
    v^2_\eta + v^2_\rho = v^2\,,
    \label{recoveringSMvev}
\end{equation}
with $v = 246$ GeV. Also, to match the 331RHN model with the SM, it can be shown that $g$ and $g_N$ are related by
\begin{equation}
    \frac{g}{g_N}=\frac{3\sqrt{2}\sin \theta_W (m_{Z^\prime}) }{\sqrt{3-4\sin^2 \theta_W  (m_{Z^\prime})}},
\end{equation}
where $\sin \theta_W(m_{Z^\prime})$ denotes the weak mixing angle evaluated at the $Z^\prime$ scale.

A notable feature of the 331RHN model is that $\sin\theta_W(m_{Z^\prime}) < \tfrac{3}{4}$. \footnote{For a detailed discussion on the value of $\sin\theta_W$ in 3-3-1 models, see Ref.~\cite{Buras:2014yna}}. Note also that $Z_\mu$ and $Z^\prime_\mu$ mix into two physical states, $Z_1 = Z \cos \theta_{331} - Z^\prime \sin \theta_{331}$ and $Z_2 = Z \sin \theta_{331} + Z^\prime \cos \theta_{331}$, where $\sin \theta_{331}$ is given by
\begin{equation}
   \sin \theta_{331} = -\frac{m_Z^2 \left(s_W^2 \left(v_\eta ^2+v_\rho ^2\right)^2- c_W^2 \left(v_\eta^2-v_\rho^2\right)^2\right)}{m_{Z^\prime}^2 \left(v_\eta^2+v_\rho^2\right)^2 \sqrt{3-4 s_W^2} }\,.
\end{equation}
This mixing angle is strongly suppressed and can be neglected in the gauge boson mass spectrum, although it plays a role in flavor-changing neutral current processes, such as meson transitions, see Refs. \cite{Buras:2014yna,Oliveira:2022dav}. 

Moreover, for $v_{\chi^\prime} \gg v_\eta, v_\rho$, it follows that
\begin{equation}
    m_{W'^+} \approx m_{U^0} \approx \frac{\sqrt{3-4\sin^2\theta_W}}{2} m_{Z^\prime} \approx 0.72\, m_{Z^\prime},
\end{equation}
implying that $Z^\prime$ is the heaviest gauge boson in the model.

The symmetric eigenstates in relations of the physical eigenstates are given by
\begin{eqnarray}
    &&W^1_\mu = \frac{W^+_\mu + W^-_\mu}{\sqrt{2}}\,\,,\,\,W^2_\mu =i \frac{W^+_\mu - W^-_\mu}{\sqrt{2}},\nonumber \\
    &&W^4_\mu = \frac{U^0_\mu + U^{0 \dagger}_\mu}{\sqrt{2}}\,\,,\,\,W^5_\mu =i \frac{U^0_\mu - U^{0 \dagger}_\mu}{\sqrt{2}},\nonumber \\
    &&W^6_\mu = \frac{W^{\prime +}_\mu + W^{\prime -}_\mu}{\sqrt{2}}\,\,,\,\,W^7_\mu =i \frac{W^{\prime +}_\mu - W^{\prime -}_\mu}{\sqrt{2}},\nonumber \\
    &&W^3_\mu= c_w Z_\mu + S_w A_\mu ,\nonumber \\
    &&W^8_\mu = \frac{t_w s_w}{\sqrt{3}}Z_\mu -\frac{s_w}{\sqrt{3}}A_\mu +\sqrt{1-t^2_w/3}Z^{\prime}_\mu,\nonumber \\
    && W^N_\mu =\sqrt{1-t^2_w/3}c_w A_\mu - \sqrt{1-t^2_w/3}s_w Z_\mu +\frac{t_w}{\sqrt{3}}Z^{\prime}_\mu.
\end{eqnarray}

The contributions of the new gauge bosons to the $S$, $T$, $U$ parameters  were analized in Ref.\cite{Long:1999bny}.  After the spontaneous symmetry breaking ${\cal G}_{331}  \to {\cal G}_{SM}$ the bileptons form a doublet $(W^{\prime +} \,,\, U^0)^T$ by $SU(2)_L$. The main conclusion from that work are that the oblique $S$ and $T$ decrease with higher masses of the bileptons, while the contribution of $Z^{\prime}$ becomes important around mass of $10$ TeV due to the mixing $Z-Z^{\prime}$. In other words, the contributions of the new gauge bosons to the oblique parameters are negligible for the new gauge bosons with mass at TeV scale. Such result motivates and justifies the analysis of the contributions  of the scalar sector  of the 331RHN to the oblique parameters.


\section{Computation of the scalar Contributions to the S, T, U parameters}
\subsection{Preliminaries}
In order to understand the contributions of the scalar sector of the 331RHN to the oblique parameters, it is recommended to write the potential in terms of   doublets of scalars.

After the spontaneous breaking of the ${\cal G}_{331}  \to {\cal G}_{SM}$  the three triplets above decouple in  the following representations by  $ {\cal G}_{SM}$:
\begin{eqnarray}
\Phi_1 = \left (
\begin{array}{c}
\eta^0 \\
\eta^- 
\end{array}
\right ),\,\Phi_2 = \left (
\begin{array}{c}
\rho^+ \\
\rho^0 
\end{array}
\right ),\,
\Phi_3 = \left (
\begin{array}{c}
\chi^0 \\
\chi^{-}
\end{array}
\right ),\, \eta^{\prime 0} ,\, \rho^{\prime +},\, R_{\chi^{\prime}}
\label{scalarcont} 
\end{eqnarray}
In the decoupling limit $ R_{\chi^{\prime}}$ does not mix neither with $\eta^0$ nor with $\rho^0$. Then it does not contribute to the oblique parameters. On the contrary, the singlets $\eta^{\prime 0}$ and $\rho^{\prime +}$ mix with the components of the doublet $\Phi_3$ and will contribute to the oblique parameters. In summary, in the decoupling limit, except by $R_{\chi^{\prime}}= H^{\prime}$, all the other scalars of the triplets will contribute to the oblique parameters.

\par In order to let this very clear, observe that the physical observables are basis independent, and it 
is often convenient to perform a basis transformation in the scalar field space.
 In particular, one may rotate to the Higg basis\cite{baseH}\cite{HiggsBase1}\cite{HiggsBase2}, where only one linear  combination of  the original doublets, $\tilde{\eta}$ and $\tilde{\rho}$ ,  carries the entire electroweak VEV, while  the orthogonal combination has a vanishing VEV.
\par Therefore,  in this basis, the doublets introduced in  (\ref{scalarcont}) are redefined as
 
\begin{equation}
\Phi_{1} =
\begin{pmatrix}
G^+_1 \\
\dfrac{1}{\sqrt{2}}(v + h + iG)
\end{pmatrix},
\qquad
\Phi_2 =
\begin{pmatrix}
h^+_1 \\
\dfrac{1}{\sqrt{2}}(H + i A)
\end{pmatrix}\,\,,\,\,\Phi_3 =
\begin{pmatrix}
h^+_2 \\
\dfrac{1}{\sqrt{2}}H^{\prime\prime}  
\end{pmatrix}.
\label{3tripl}
\end{equation}
\par The use of this basis is particularly well motivated when the goal is to 
systematically determine the contributions of scalar fields to gauge or matter 
currents. As a consequence, the couplings of 
the scalar sector to the gauge currents acquire a transparent structure: the 
longitudinal components of the electroweak gauge bosons couple exclusively to the 
fields contained in $\Phi_1$, while the additional scalar degrees of freedom encoded 
in $\Phi_2$, $\Phi_3$ contribute only through interactions involving physical scalar states.

The potential scalar involving only  these three triplets is composed by the following terms

\begin{align}
   V_{\text{eff}}(\Phi_1,\Phi_2, \Phi_3 ) \;=&\;
\mu_1^2\, \Phi_1^\dagger \Phi_1  +\mu_2^2\, \Phi_2^\dagger \Phi_2 
+ \mu_{12}^2\, \Phi_1^\dagger  \Phi_2  + \mu_3^2\, \Phi_3^\dagger \Phi_3 
+ \lambda_1 \left(\Phi_2^\dagger \Phi_2\right)^2
+ \lambda_2  \left(\Phi_1^\dagger \Phi_1\right)^2
\nonumber\\[4pt]
&  + \lambda_3  \left(\Phi_3^\dagger \Phi_3\right)^2
+ \lambda_{12}\, (\Phi_2^\dagger \Phi_2)(\Phi_1^\dagger \Phi_1)
+ \zeta_{12}\, (\Phi_1 \Phi_2)^\dagger
(\Phi_1 \Phi_2)
\nonumber\\[4pt]
& + \lambda_{13}\,(\Phi_1^\dagger \Phi_1)(\Phi_3^\dagger \Phi_3)
+\lambda_{23}\,(\Phi_2^\dagger \Phi_2)(\Phi_3^\dagger \Phi_3)
\nonumber\\
&+\lambda'_{13}\,(\Phi_1^\dagger \Phi_3)(\Phi_3^\dagger \Phi_1)
+\lambda'_{23}\,(\Phi_2^\dagger \Phi_3)(\Phi_3^\dagger \Phi_2)
\nonumber\\
& +\frac{\lambda''_{12}}{2}
\Big[(\Phi_1^\dagger \Phi_2)^2+\text{h.c.}\Big]
+\frac{\lambda''_{13}}{2}
\Big[(\Phi_1^\dagger \Phi_3)^2+\text{h.c.}\Big]
+\frac{\lambda''_{23}}{2}
\Big[(\Phi_2^\dagger \Phi_3)^2+\text{h.c.}\Big]. 
\label{331effp}
\end{align}

\subsection{Scalar contributions to $\Pi_{WW}$  and  $\Pi_{ZZ}$ vacuum polarizations from  $331RHN$ model}

\par In the $331RHN$ model, the presence of additional scalar degrees of freedom modifies the electroweak gauge-boson propagators through loop-induced corrections. These effects are conveniently described in terms of vacuum polarization functions, which can be expressed as correlators of the weak isospin currents.
\par In particular, the contributions associated with the charged and neutral weak currents are given by
\begin{equation}
\Pi^{11}_{\mu\nu}(q)
=
 i\int d^4x\, e^{iq\cdot x}
 \langle 0| T\{ J_\mu^1(x) J_\nu^1(0) \} |0\rangle = \left(q^2g_{\mu\nu} -  q_\mu q_\nu \right)\Pi_{11}(q^2),
 \label{eqpol11}
\end{equation}
and
\begin{equation}
\Pi^{33}_{\mu\nu}(q)
=
 i\int d^4x\, e^{iq\cdot x}
 \langle 0| T\{ J_\mu^3(x) J_\nu^3(0) \} |0\rangle  = \left(q^2g_{\mu\nu} -  q_\mu q_\nu \right)\Pi_{33}(q^2),
 \label{eqpol33}
\end{equation}
where $J_\mu^1$ and $J_\mu^3$ denote the isospin currents associated to $SU(2)_L$ generators, $\sigma^1$ and $\sigma^3$.  
\par In the above expression,  the superscripts $(11)$ and $(33)$ label the scalar components
of the polarizations of the gauge bosons  in the custodial $SU(2)$ basis.

\par The quantity $\Pi^{11}_{\mu\nu}$ corresponds to the vacuum polarization associated with the charged isospin current, while $\Pi^{33}_{\mu\nu}$ is associated with the neutral isospin current. These correlators are directly related to the gauge-boson self-energies, such that the transverse parts of $\Pi^{11}_{\mu\nu}(q)$ and $\Pi^{33}_{\mu\nu}(q)$ determine, respectively, the $W$ and $Z$ boson vacuum polarizations, denoted by $\Pi_{WW}(q^2)$ and $\Pi_{ZZ}(q^2)$,  evaluated at zero momentum transfer $q^2 \to 0$. 
\par  The explicit form of these currents follows from the kinetic terms of the scalar doublets,
\begin{equation}
\mathcal{L}_{\rm kin} = \sum^{3}_{i=1} (D_\mu \Phi_i)^\dagger (D^\mu \Phi_i),
\label{eqcurrc}
\end{equation}
through the Noether procedure. As a result, the currents are bilinear in the scalar fields and involve derivative couplings between charged and neutral components. After rotating to the mass eigenstate basis, these currents generate interactions of the form (schematically)
\begin{equation}
J_\mu \supset  
H\partial_\mu h^+_1
-
h^+_1\partial_\mu H
+\text{h.c.},
\end{equation}
which connect pairs of charged and neutral scalars.
\begin{figure}[t]
\vspace*{-0.45cm}
    \centering
\hspace*{-0.2cm}\includegraphics[scale=0.7]{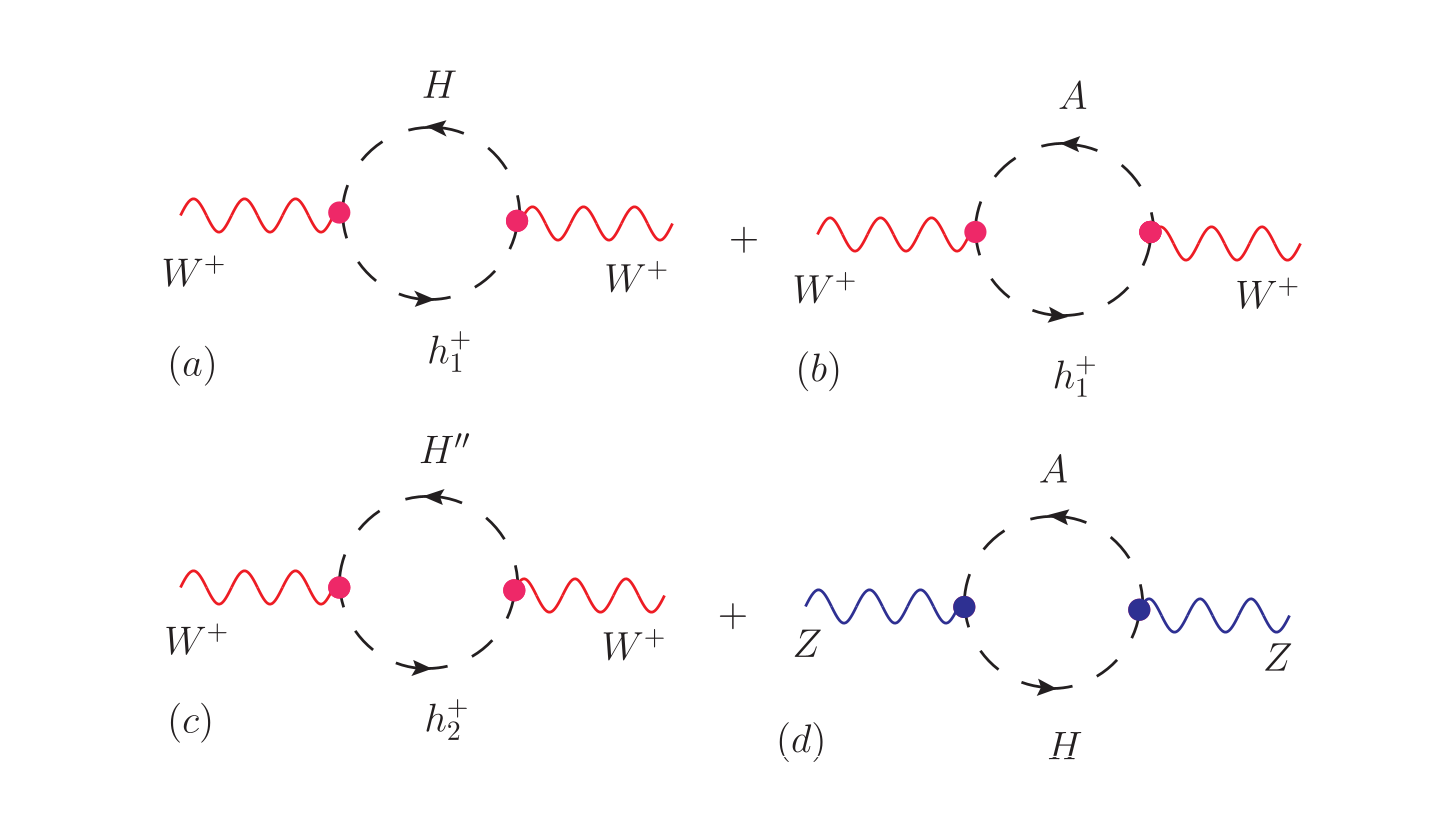}
    \caption{Feynman diagrams  at  one-loop contributing  for  the vacuum polarization functions defined in Eqs.(\ref{pol11c}) and (\ref{pol11cb}). }
    \label{fig4}
\end{figure}
\par  Using the scalar doublets given in Eq.~(\ref{3tripl}), the corresponding weak isospin currents entering Eqs.~(\ref{eqpol11}) and (\ref{eqpol33}) can then be explicitly constructed and are listed below
\begin{align}
J_\mu^{1} & =  \frac{i}{2\sqrt{2}}\Big( h^+_1\hat{\partial}_\mu H
-H\hat{\partial}_\mu h^+_1 \,  +  A\hat{\partial}_\mu h^+_1
- h^+_1\hat{\partial}_\mu A \Big) + H.c. \nonumber \\
J_\mu^{3} &= 
\frac{i}{2\sqrt{2}}\Big(  H\hat{\partial}_\mu A - H\hat{\partial}_\mu A \Big) + H.c.,
\label{331currA}
\end{align}
\noindent where  we define 
\begin{align}
\hat{\partial}_\mu h^+_1
&=
\partial_\mu h^+_1
- i g
\left(
\frac{1}{2} W_\mu^3h^+_1
+ W_\mu^+ H
\right)\label{partplus}\\
\hat{\partial}_\mu H
&=
\partial_\mu H
- i g
\left(
W_\mu^- h^+_1
- \frac{1}{2} W_\mu^3 H
\right)\label{partH} \\
\hat{\partial}_\mu A
&=
\partial_\mu A
- i g
\left(
W_\mu^- h^+_1
- \frac{1}{2} W_\mu^3 A
\right)\label{partA}.
\end{align} 
\par In addition to the doublets $\Phi_1$ and  $\Phi_2$, whose contributions were listed above, we have the third scalar doublet $\Phi_3$, which effectively adds a contribution to Eq.(\ref{eqpol11}) due to the current 
\begin{align}
J_\mu^{1^\prime} =  \frac{i}{2\sqrt{2}}\left(H^{\prime\prime} \hat{\partial}_\mu h^+_2
- h^+_2\hat{\partial}_\mu H^{\prime\prime} \right) + H.c, 
\label{331newcurr}
\end{align}
\noindent where in this case the expressions for  $(\hat{\partial}_\mu h^+_2 , \hat{\partial}_\mu H^{\prime\prime})$ are obtained from the substitutions $(h^+_1\to h^+_2)$ and  $(H\to H^{\prime\prime})$ in the Eqs.(\ref{partplus}) and (\ref{partH}) .
\par From the explicit form of the scalar currents presented in Eqs.(\ref{331currA}) and (\ref{331newcurr}), one can directly identify the interaction vertices that couple pairs of scalar fields through derivative structures. These vertices define, in a straightforward manner, the set of one-loop Feynman diagrams in which charged and neutral scalars propagate inside the loop and contribute to the vacuum polarization functions. 
\par As a result, we  identify the corrections to the vacuum polarization functions defined in Eqs. (\ref{eqpol11}) and (\ref{eqpol33}) as 
\begin{align}  
\Pi^{11}_{\mu\nu}(q)  & =  i\int d^4x\, e^{iq\cdot x}\Big(\langle J^1_\mu(x)J^1_\nu(0)\rangle_{ Hh^+_1} + \langle J^1_\mu(x)J^1_\nu(0)\rangle_{ Ah^+_1} + \langle J^1_\mu(x)J^1_\nu(0)\rangle_{H^{\prime\prime} h^+_2}\Big)\label{pol11c}\\ \nonumber\\
\Pi^{33}_{\mu\nu}(q)  & = i\int d^4x\, e^{iq\cdot x}\ \langle J^3_\mu(x)J^3_\nu(0)\rangle_{AH} \label{pol11cb}, 
\end{align}
where in the compact notation used, the subscript indices label the respective corrections and the diagrammatic interpretation can be established through the Feynman diagrams shown in Fig.1. 
\par The corrections identified in the labels of the gauge boson correlators will be evaluated in the next section. In particular, these contributions induce modifications to the oblique parameters $S$, $T$, and $U$. By computing the corresponding vacuum polarization functions and extracting their momentum-dependent behavior, we  will determine the impact of the scalar states on the electroweak oblique corrections.

\subsection{Scalar contributions  from  $331RHN$ model to  oblique parameters}

\par The effects of new scalar states on electroweak precision observables
can be  encoded in the oblique parameters $S$, $T$, and $U$,
which are defined in terms of the transverse parts of the gauge boson
vacuum polarization functions. In the $ SU(2)_L \times U(1)_Y $ basis,
it is convenient to express these quantities in terms of the correlators
$\Pi_{11}(q^2)$ and $\Pi_{33}(q^2)$, which we defined in Eqs.(\ref{eqpol11}) and (\ref{eqpol33}).

In terms of these correlators, the oblique parameters can be written as\cite{PeskinTakeuchi1990,PeskinTakeuchi1992}
(see also Ref.~\cite{AltarelliBarbieri1991})
\begin{align}
\alpha S &= 4 s_W^2 c_W^2 
\left[
\Pi'_{33}(0) - \Pi'_{3Q}(0)
\right] \\
\alpha T &= 
\frac{\Pi_{11}(0) - \Pi_{33}(0)}{m_W^2} \\
\alpha U &= 4 s_W^2 
\left[
\Pi'_{11}(0) - \Pi'_{33}(0)
\right],
\end{align}
where the prime denotes differentiation with respect to $q^2$
evaluated at $q^2=0$, $\alpha = \frac{e^2}{4\pi}$ and $e^2 = g^2s_W^2$.
\par In the expression above, the mixed vacuum polarization function
$\Pi_{3Q}(q^2)$ is defined through the correlator between the third
component of the weak isospin current, $J_\mu^3$, and the
electromagnetic current, $J_\nu^{Q}$. 

\par Among these parameters, $T$ is particularly sensitive to new scalar
contributions. This is because $T$ measures the breaking of custodial
symmetry through the difference between neutral and charged current
correlators at zero momentum. Mass splittings within scalar doublets,
such as those arising in $331RHN$ model, directly induce
non-vanishing contributions to $\Pi_{11}(0) - \Pi_{33}(0)$.

\par The scalar vacuum polarization functions $\Pi_{11}(q^2)$ and
$\Pi_{33}(q^2)$ are extracted from the transverse part of the
current--current correlators in the  Eqs.(\ref{eqpol11}) and (\ref{eqpol33}), 
where  $\Pi_{11}(q^2)$ and $\Pi_{33}(q^2)$ are computed from one-loop Feynman
diagrams depicted in the Fig.1. 

\par  The corresponding amplitudes are evaluated at general external momentum $q$,
and the respective  coefficients of the transverse tensor structure are then identified.  
Finally, with these results in hand,  we determine the corrections to the scalar function
$\Pi_{11}(q^2)$ after factoring out $(q^2 g_{\mu\nu} - q_\mu q_\nu)$ from the full amplitude and 
the scalar components  $\Pi_{11}(0)$ are obtained  by taking the limit $q^2 \rightarrow 0$.

\par  An analogous procedure applies to $\Pi_{33}(0)$ and $\Pi^\prime_{3Q}(q^2)|_{{}_{q^2 \to 0}}$, and the scalar contributions of the  $331RHN$ model to  oblique parameters can be summarized as 
\begin{align}
\alpha S  &=
\frac{e^2}{192\pi^2s^2_W}\Big( G(m^2_{H},m^2_A) - \frac{1}{2}\log[\frac{m^2_{h^+_1}}{m^2_H}]  - \frac{1}{2}\log[\frac{m^2_{h^+_1}}{m^2_A}] + \log\Big(\frac{m^2_{h^+_2}}{m^2_{H^{\prime\prime}}}\Big)\Big) \label{parS} \\
 \alpha T  &= \frac{e^2}{64\pi^2s^2_Wm^2_W}\Big(F(m^2_{h^+_1},m^2_H) + F(m^2_{h^+_1},m^2_A) + F(m^2_{h^+_2},m^2_{H^{\prime\prime}})  - F(m^2_{H},m^2_A)\Big) \label{parT} \\
\alpha U  &= \frac{e^2}{192\pi^2s^2_W}\Big(G(m^2_{h^+_1},m^2_H) + G(m^2_{h^+_1},m^2_A) + G(m^2_{h^+_2},m^2_{H^{\prime\prime}}) - G(m^2_{H},m^2_A)\Big) ,
\label{parU}
\end{align}
\noindent where the  functions  $F(m^2_{1},m^2_2)$ and  $G(m^2_{1},m^2_2)$  are given by  
\begin{align}
     F(m^2_{1},m^2_2) & = 
\left[
\frac{m_{1}^2+m_{2}^2}{2}
-
\frac{m_{1}^2 m_{2}^2}{m_{1}^2-m_{2}^2}
\ln\!\left(\frac{m_{1}^2}{m_{2}^2}\right) 
\right] ,\nonumber \\
 G(m^2_{1},m^2_2) & = -\frac{5}{6} + \frac{m^2_{1}m^2_{2}}{(m^2_{1}-m^2_{2})^2}
    +  \frac{m^4_{1}(m^2_{1} - 3m^2_{2})}{(m^2_{1} -m^2_{2})^3}\log[\frac{m^2_{1}}{m^2_2}]. 
    \label{funcFeG}
\end{align}

\section{NUMERICAL RESULTS}

\par In this section, we analyze the contributions of the scalar sector defined in Eq.~(\ref{3tripl}) to the electroweak oblique parameters $S$, $T$, and $U$, introduced in Eqs.~(\ref{parS})-(\ref{parU}).  Assuming the CP-even scalar mass matrix defined in Eq.~(\ref{masseven2}), and working in the Higgs basis,  the diagonalization procedure yields the following mass to the scalars $h$ and $H$
\begin{align}
m_h^2 &= \frac{v^2(\lambda_2\lambda_3 - \lambda^2_6)}{\lambda_2 + \lambda_3 -2\lambda_6}, \label{hmass}\\
m_H^2 &= m^2_A + v^2\frac{(\lambda_2 - \lambda_6)(\lambda_3 - \lambda_6)}{\lambda_2 + \lambda_3 -2\lambda_6} \label{Hmass}.
\end{align}
\par Here, $h$ is identified with the SM-like Higgs boson, while $H$ corresponds to the additional CP-even neutral scalar state.  The mass of the CP-odd pseudoscalar  is given by
\begin{align}
m^2_A &= \frac{f v_{\chi^{\prime}}}{4}\left( \frac{v_\eta v_\rho}{v^2_{\chi^{\prime}}} + \frac{v_\eta}{v_\rho} + \frac{v_\rho}{v_\eta}\right) \nonumber \\
&= \frac{f v_{\chi^{\prime}}}{4}\tan\phi \left( 1 + \frac{1}{\tan^2\phi }\right).
\label{Amass}
\end{align}
 
\par In addition to the CP-even states $h$ and $H$, the neutral scalar spectrum also contains a heavier state $H^{\prime\prime}$
whose mass  corresponds to
\begin{align}
m^2_{H^{\prime \prime}} &=
\frac{\lambda_7}{4}\left(v_{\chi^{\prime}}^2+ v^2_\eta\right)
+ \frac{f v_\rho}{4}\left(\frac{v_{\chi^{\prime}}}{v_\eta}
+ \frac{v_\eta}{v_{\chi^{\prime}}}\right) \nonumber  \\
& \approx m^2_A + \frac{\lambda_7 v_{\chi^{\prime}}^2}{4},
\label{Hppmass}
\end{align}
\par The scalar mass spectrum becomes complete once the charged scalar states 
$h_1^+$ and $h_2^+$, listed previously are included. Their masses can be written as 
\begin{align}
m_{h_1^+}^2 
&= \left( \frac{f v_{\chi^{\prime}}}{2}  
+ \frac{\lambda_9}{2}v_\eta v_\rho \right)
\left(\frac{v_\eta}{v_\rho} + \frac{v_\rho}{v_\eta} \right) \nonumber \\
&= 2m^2_A + \frac{\lambda_9 v^2}{2}, \label{Hc1mass}\\
m_{h_2^+}^2 
&= \left(\frac{f v_{\eta}}{2}+ \frac{\lambda_8}{2} v_\rho v_{\chi^{\prime}}\right)
\left(\frac{v_{\chi^{\prime}}}{v_\rho}
+\frac{v_\rho}{v_{\chi^{\prime}}}\right) \nonumber \\
& \approx \frac{2m^2_A}{\tan^2\phi} + \frac{\lambda_8}{2} v^2_{\chi^{\prime}}.
\label{Hc2mass}
\end{align}

\par  With the exception of the SM-like Higgs boson mass $m_h$, one can observe from 
Eqs.~(\ref{Hmass})-(\ref{Hc2mass}) that all remaining scalar masses depend 
explicitly on combinations of the parameters $f$ and $v_{\chi^{\prime}}$. 

\par This structure implies that the heavy scalar spectrum is governed by 
the interplay between $f$ and $v_{\chi^{\prime}}$, leading to correlated mass 
patterns among the CP-even, CP-odd, and charged states. In particular, 
increasing $v_{\chi^{\prime}}$  drives the masses of $H$, $A$, 
$H^{\prime\prime}$, and $h_i^+$ to higher values, signaling a decoupling 
behavior from low-energy electroweak observables.

\par  Conversely, moderate values  of $f$ and $v_{\chi^{\prime}}$ may induce sizable mass splittings within the 
scalar doublets , potentially generating significant contributions to the 
oblique parameters, especially to the parameter $T$ which is directly sensitive to custodial symmetry breaking and receives enhanced contributions when sizable mass splittings occur between $(H, H^{\prime\prime},A, h_1^+,h_2^+)$ states. 

\par The current global fits to electroweak precision data constrain the oblique parameters to lie close to their Standard Model predictions. In particular,  the most recent determination of the $T$ parameter yields a value consistent with\cite{PDG}
\begin{align}
T = 0.01 \pm 0.12. 
\label{Tlim}
\end{align}
\par  By construction, non-vanishing contributions to the oblique parameters 
originate from sectors beyond the standard model.  Within the framework of the $331RHN$ model, the additional quarks 
$(u'_3, d'_i)$ appearing in the triplet representations of Eq.~(\ref{quarks}) 
are singlets under $SU(2)_L$. As a consequence, they do not induce 
direct contributions to the electroweak oblique parameters.  

\par Regarding the gauge sector, the contributions from the $(W^\prime, Z^\prime, U^0)$ gauge bosons 
 were previously evaluated in Ref.~\cite{Long:1999bny}.  It was shown that these corrections are strongly suppressed and therefore  remain negligible when compared to the current experimental precision  on the $S$, $T$, and $U$ parameters.

\par To identify the regions compatible with the electroweak constraint on the 
$T$ parameter, we perform parameter scans in the multidimensional space 
$(f, v_{\chi^{\prime}}, \lambda_2, \lambda_6, \lambda_7, \lambda_8, \lambda_9)$, where we assume $\lambda_2 =\lambda_3$. 
For the numerical analysis, we generate random points within the ranges
\begin{align}
& 0.01 \leq f \leq 50, \nonumber \\
& 500~\text{GeV} \leq v_{\chi^\prime} \leq 16~\text{TeV}, \nonumber \\
& 0 < \lambda_2, \lambda_7, \lambda_8, \lambda_9  \leq 5, \nonumber \\
& 0 < \tan\phi < 60,
\end{align}
while fixing the SM-like Higgs mass to $m_h = 125.1~\text{GeV}$ through the relation 
\[
\lambda_2 + \lambda_6 = \frac{2m_h^2}{v^2}.
\]
\par The results of the scan are shown in Fig.~2. The upper panels display the 
values of $m_H$, whereas the lower panels correspond to the charged scalar 
mass $m_{h_1^\pm}$ in the $(f, v_{\chi^\prime})$ plane.
\begin{figure}[t]
    \centering
\includegraphics[scale=0.4]{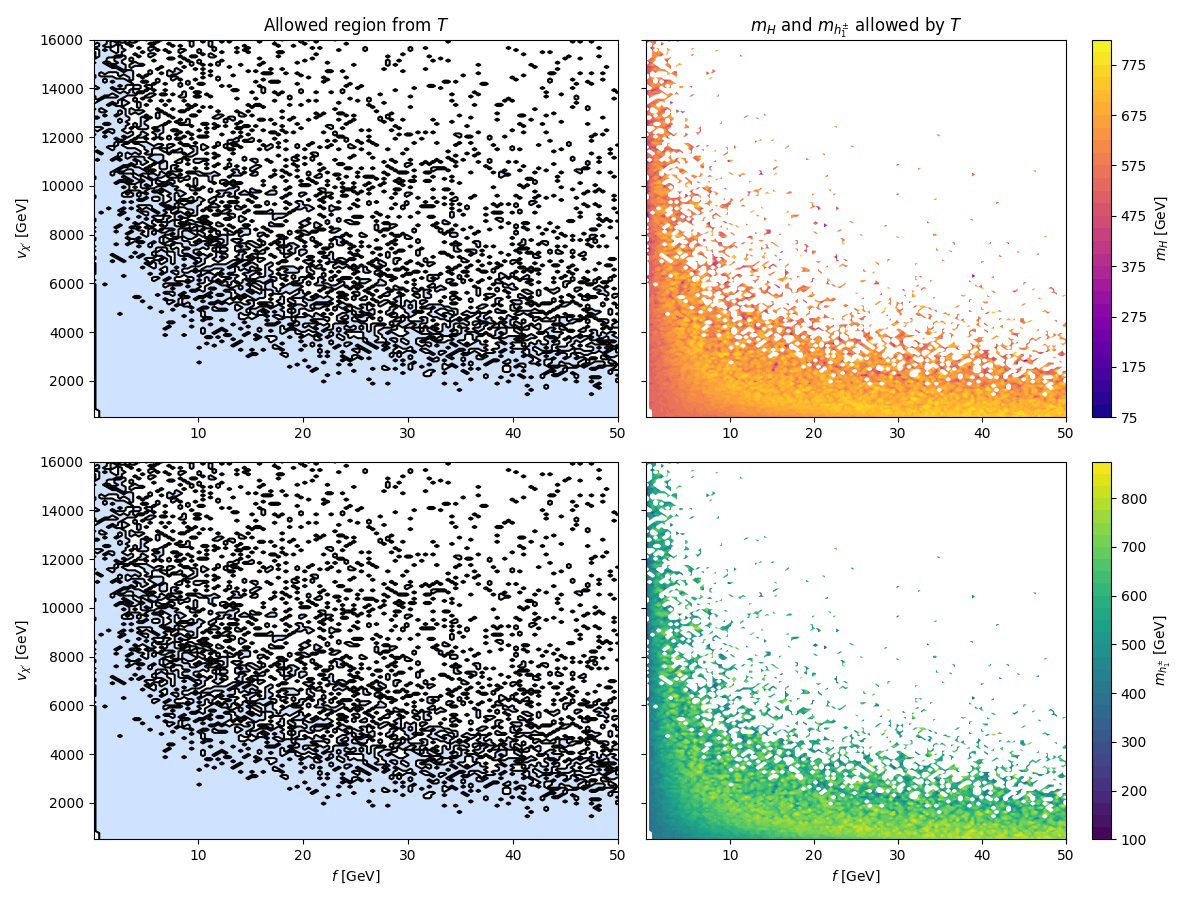}
    \caption{
Predictions of the $331RHN$ model for the oblique parameter $T$ in the 
$(f, v_{\chi^{\prime}},..\lambda_9)$ plane. The left panel show the results 
to $T$ as function of  $(f, v_{\chi^{\prime}},..\lambda_9)$. In the right 
panel we show  $m_H$ and $m_{h_1^\pm}$ behavior, where the white distribution 
corresponds to the region where  the oblique parameter reaches its experimental upper limit.}
\end{figure}

\par  In each figure, the left panel shows the region compatible with the $T$ 
constraint (light blue area), while the right panel presents the corresponding 
scalar mass spectrum.  The upper boundary of the white distribution corresponds to
the region where  the oblique parameter reaches its experimental upper limit.
The region below this boundary defines the parameter space compatible with the $T$ constraint.

\par The color bar represents the magnitude of the scalar masses over the 
scanned parameter space. From the upper right panel, one observes that 
the $T$ constraint restricts the charged scalar mass to 
$m_{h_1^\pm} \lesssim 800~\text{GeV}$. 
Analogously, the lower right panel shows that the neutral scalar mass 
is bounded by $m_H \lesssim 770~\text{GeV}$ once the $T$ limit is imposed.

\par The sizable white regions visible in both panels highlight the strong 
restrictive power of the $T$ parameter, illustrating the substantial 
reduction of the viable scalar mass spectrum due to electroweak 
precision constraints.

\par  To quantify the impact of the inert doublet $\Phi_3$ on the electroweak 
constraint encoded in the $T$ parameter, we analyze in Fig.~3 the regime 
in which the scalar potential exhibits a decoupling pattern, namely 
$\lambda_7 = \lambda_8 \ll \lambda_{2,6,9}$. 
In this limit, the scalar states originating from $\Phi_3$ become 
effectively decoupled from the low-energy spectrum, so that their 
contributions to $T$ are  suppressed. 
\par  This setup allows us to isolate the residual effects on the masses of 
$H$ and $h_1^\pm$, as well as on the bounds for the symmetry-breaking 
scale $v_{\chi^{\prime}}$, in the absence of sizable inert-doublet 
corrections.

\begin{figure}[t]
    \centering
\includegraphics[scale=0.4]{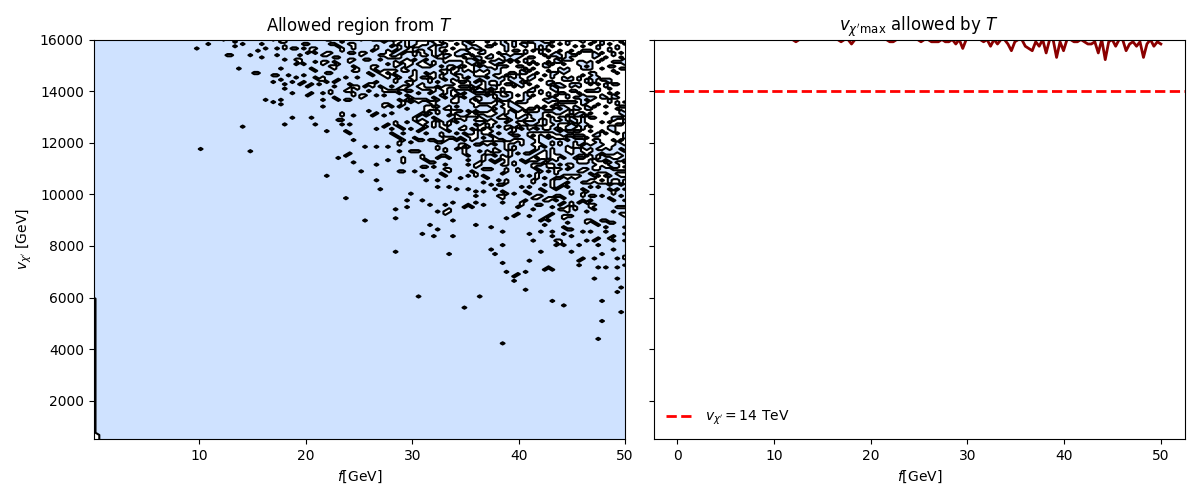}
\includegraphics[scale=0.4]{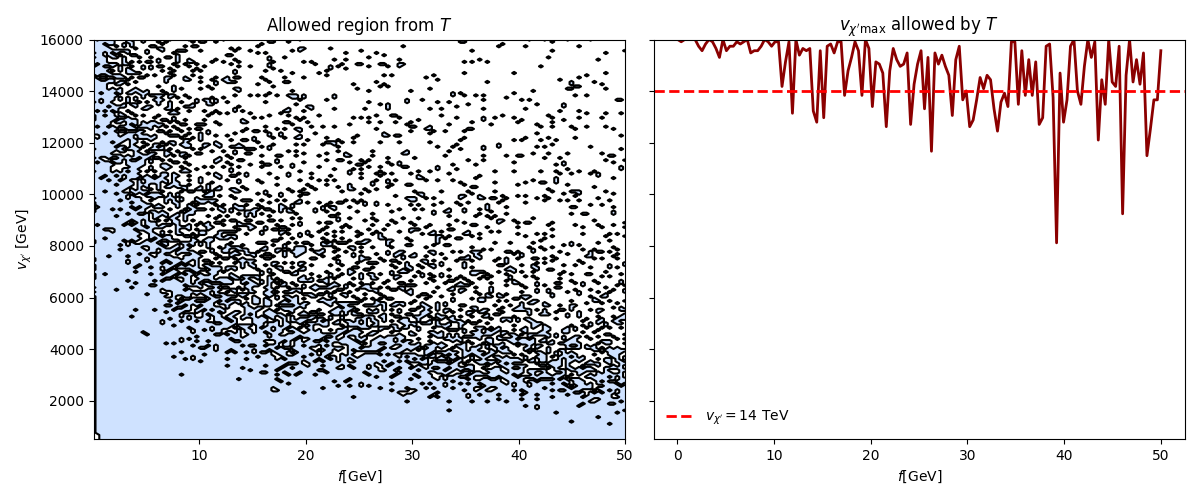}
   \caption{
Left panel: behavior of $T$ in the $(f, v_{\chi^{\prime}}, \ldots, \lambda_9)$ space after imposing the experimental constraint; right panel: maximum allowed $v_{\chi^{\prime}}(T)$ from the full scan, assuming $\lambda_7=\lambda_8=10^{-3}$ (upper) and $\lambda_7=\lambda_8\sim\lambda_{2,6,9}$ (lower).
}
\end{figure}

\par In the upper panels of Fig.~3, the left plot shows the behavior of the 
$T$ parameter as a function of  $(f, v_{\chi^{\prime}}, \ldots, \lambda_9)$, 
after imposing the experimental $T$ constraint. 
The right plot displays the maximum allowed value of $v_{\chi^{\prime}}$ 
obtained from the scan over the full parameter space under the same 
decoupling assumption. 
\par  Since in this regime the heavy scalars originating from $\Phi_3$ do not 
provide sizable contributions to $T$, no significant upper bound on 
$v_{\chi^{\prime}}$ is observed in the high-scale region of the figure.

\par In contrast, the lower panels correspond to the non-decoupled scenario, 
where $\lambda_7 = \lambda_8 \sim \lambda_{2,6,9}$.  In this case, the contributions
from $m_{H^{\prime\prime}}$ and  $m_{h_2^\pm}$ are fully restored and generate additional corrections 
to the $T$ parameter.   

\par The lower right panel shows that the symmetry-breaking scale becomes 
bounded by  $v_{\chi^{\prime}} \lesssim 14~\text{TeV}$,   in order to satisfy the electroweak precision constraint on the $T$ 
parameter. This behavior originates from the positive contribution of 
the inert doublet $\Phi_3$ to $T$, induced by the loop function  $F(m^2_{h_2^+}, m^2_{H^{\prime\prime}})$.

\par To clarify this point, let us consider, for simplicity, 
the approximate limit $m_A \simeq m_H$, such that  $F(m_H^2, m_A^2) \approx 0$. 
In the decoupling regime $\lambda_7=\lambda_8 \ll \lambda_{2,6,9}$, 
the dominant contribution to $T$ can be approximated as

\begin{align}
T &\simeq \frac{1}{16\pi s_W^2 M_W^2}
\,2F(m_{h_1^+}^2, m_A^2) \nonumber \\
&\simeq \frac{3}{64\pi s_W^2 M_W^2}
\left( f v_{\chi^{\prime}} \tan\phi + 2\lambda_9 v^2 \right).
\end{align}

\par Defining
\[
\Lambda^2 \equiv f v_{\chi^{\prime}} \tan\phi + 2\lambda_9 v^2,
\qquad
A \equiv \frac{64\pi s_W^2 M_W^2}{3},
\]
the condition $T=T_{\max}$ implies
\[
\Lambda^2_{\max} = A\,T_{\max}.
\]
\par This relation defines,  in the $(f- v_{\chi^{\prime}})$ spacee,  the 
maximal region compatible with the $T$ constraint. 
As shown in the upper left panel of Fig.~3, this region is essentially 
saturated for $f \sim (1\text{--}10)~\text{GeV}$, indicating that, in the 
decoupled scenario, the $T$ parameter does not impose a significant 
restriction on $v_{\chi^{\prime}}$. 
In this regime, one effectively obtains 
$\Lambda^2_{\max} \sim fv_{\chi^{\prime}}$, 
 and therefore no meaningful upper constraint on the 
symmetry-breaking scale arises.

\par Once the inert-doublet interactions are restored, assuming 
$\lambda_7=\lambda_8 \sim \lambda_{2,6,9}$, the $T$ parameter becomes

\begin{align}
T &\simeq \frac{1}{16\pi s_W^2 M_W^2}
\left[2F(m_{h_1^+}^2, m_A^2) 
+ F(m_{h_2^+}^2, m_{H^{\prime\prime}}^2)\right] \nonumber \\
&\simeq \frac{3}{64\pi s_W^2 M_W^2}
\left( f v_{\chi^{\prime}} \tan\phi 
+ 2\lambda_9 v^2 
+ 2\lambda v_{\chi^{\prime}}^2 \right),
\end{align}
where we have defined $\lambda=\lambda_7=\lambda_8$.

\par In this case, the maximal allowed region becomes
\[
\Lambda_{\max}^{2\,\prime}
=
A\,T_{\max}
-
2\lambda v_{\chi^{\prime}}^2,
\]
which is manifestly smaller than $\Lambda_{\max}^2$. 
The additional positive contribution proportional to 
$\lambda v_{\chi^{\prime}}^2$ further restricts the viable parameter space, 
as illustrated in the lower left panel of Fig.~3.

\par Consequently, even within the same interval 
$f \sim (1\text{--}10)~\text{GeV}$, once 
$\Lambda_{\max}^{2\,\prime}$ defines the region allowed by the $T$ bound and satisfies 
$\Lambda_{\max}^{2\,\prime} < \Lambda_{\max}^2$, the inclusion of $\Phi_3$ 
 reduces the maximal value of the symmetry-breaking scale. 
This implies $v_{\chi^{\prime}}(\lambda) < v_{\chi^{\prime}}(0)$, where 
the notation $\lambda$ denotes the scenario in which the inert-doublet 
interactions are restored.

\par As illustrated in the lower right panel of Fig.~3, this behavior translates into 
an upper bound $v_{\chi^{\prime}} \lesssim \mathcal{O}(14)~\text{TeV}$. 
In addition, the parameter $f$ also becomes constrained from above by the 
$T$ bound. The scan leads to the scenario shown in the lower 
left panel, where $f_{\max} \lesssim \mathcal{O}(10)~\text{GeV}$. 
This behavior reflects the enhanced custodial-symmetry breaking induced 
by the heavy scalar states $m_{H^{\prime\prime}}$ and $m_{h_2^\pm}$.

\par At this point, it is worth noting that the parameter 
$f$ controls the mass splitting 
among the scalar states and therefore plays a crucial role in determining 
the magnitude of the custodial-symmetry breaking effects entering the 
$T$ parameter. Larger values of $f$ enhance the scalar mass differences 
and consequently increase the contribution to $T$.

\par Recently, within the framework of the minimal 331 model, the authors of 
Ref.~\cite{STU-MIN} analyzed the indirect effects of the scalar sector on the 
electroweak oblique parameters $S$, $T$, and $U$ and as a result verified the existence 
of a constrain  on vacuum expectation value  $ 1500 GeV\lesssim v_3 \lesssim 2300 GeV$ 
due to the triplet $\chi$. 

\begin{figure}[t]
    \centering
\includegraphics[scale=0.4]{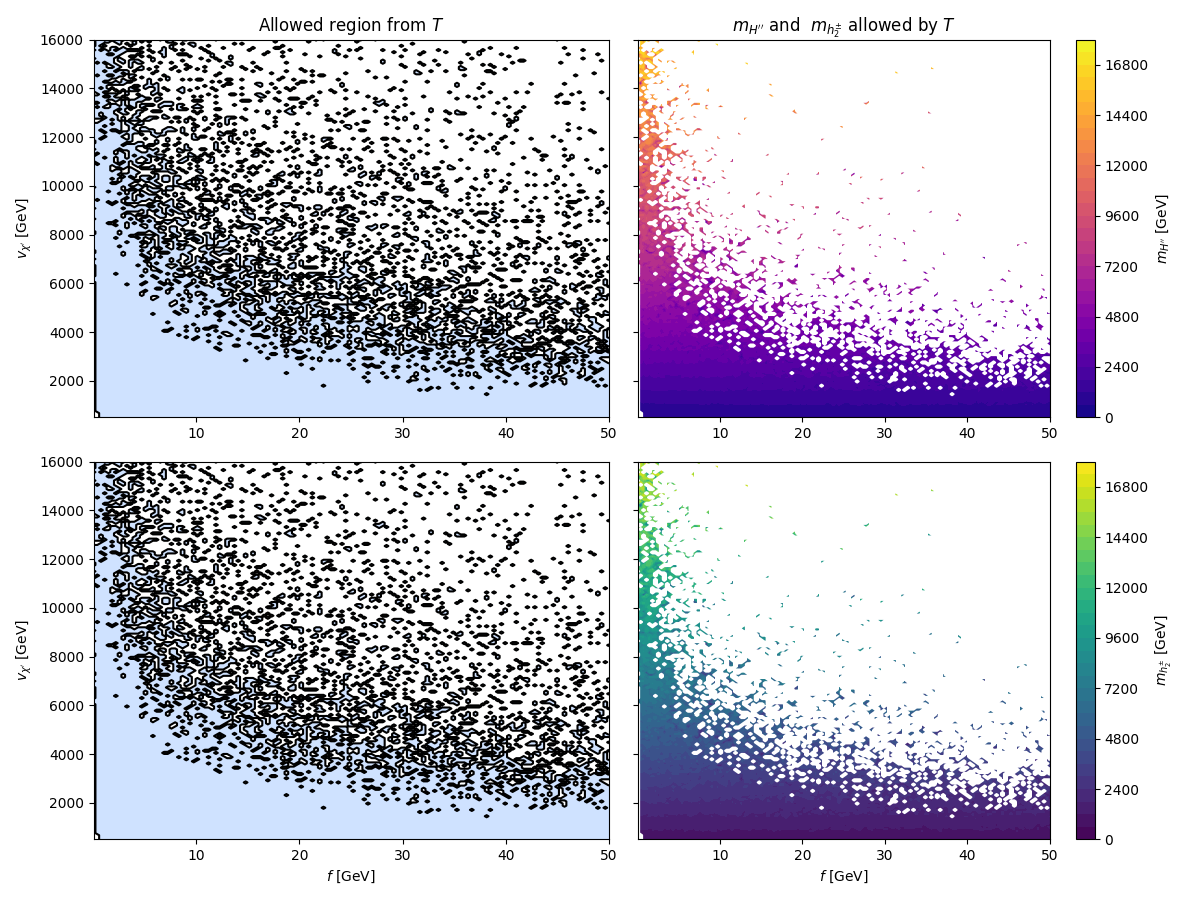}
    \caption{
Predictions of the $331RHN$ model for the oblique parameter $T$ in the 
$(f, v_{\chi^{\prime}},..\lambda_9)$ plane to $m_{H^{\prime\prime}}$ and $m_{h_2^\pm}$. The left panel show the results  to $T$ as function of  $(f, v_{\chi^{\prime}},..\lambda_9)$. In the right 
panel the white distribution corresponds to the region where  the oblique parameter
reaches its experimental upper limit.}
\end{figure}

\par The minimal 331 model features a  particle content, including 
new charged gauge bosons $U^{\pm\pm}$ and $V^{\pm}$, as well as additional 
scalar fields carrying single and double electric charges. The minimal model 
exhibits a broader scalar spectrum compared to the $331RHN$ model.
 As a consequence, stronger constraints on the symmetry-breaking scale 
$v_{\chi^\prime} \equiv v_3$ were obtained due to the cumulative effect 
of these additional scalar contributions.

\par Our results are consistent with those reported in Ref.~\cite{STU-MIN}, 
where it was found that $f_{\max} \lesssim 9~\text{GeV}$. 
This agreement indicates that electroweak precision observables impose 
nontrivial constraints on the 3-3-1 symmetry-breaking scale, mainly due to 
the contributions of the heavy scalar states generated in this symmetry-breaking sector.

\par In addition to the scalar mass behavior shown in Fig.~2, we present in  Fig.~4 
the corresponding scan results for the heavier states  $m_{H^{\prime\prime}}$ and $m_{h_2^\pm}$, obtained under the same  parameter prescription. 

\begin{figure}[t]
    \centering
\includegraphics[scale=0.4]{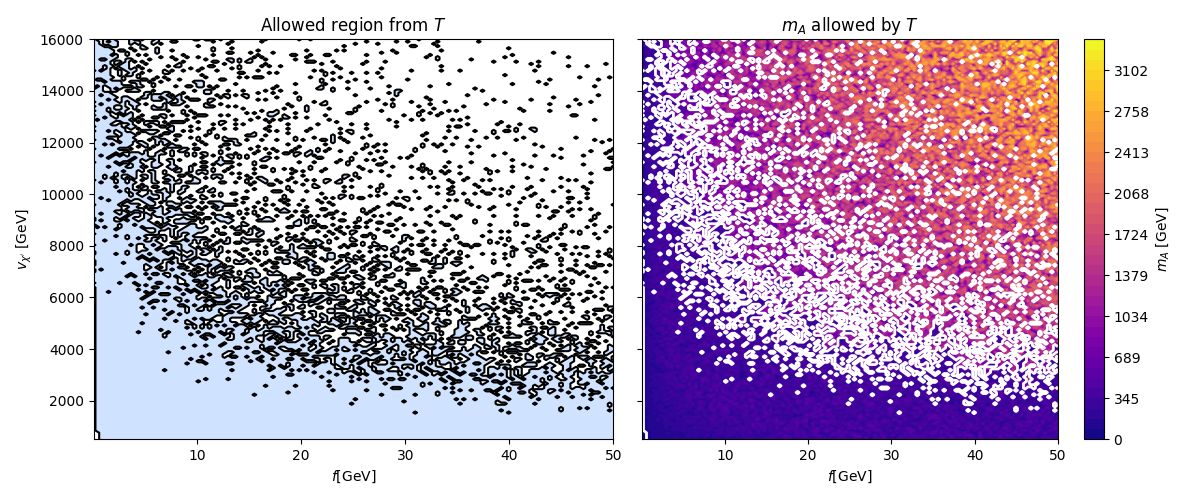}
    \caption{
Predictions of the $331RHN$ model for the oblique parameter $T$ in the 
$(f, v_{\chi^{\prime}},..\lambda_9)$ plane to $m_{A}$ . The left panel show the results  to $T$ as function of  $(f, v_{\chi^{\prime}},..\lambda_9)$. In the right  panel the white distribution corresponds to the region where  the oblique parameter reaches its experimental upper limit.}
\end{figure}

\par Although the scan permits values as large as  $v_{\chi^\prime} \simeq 16~\text{TeV}$, the imposition of the $T$  constraint effectively reduces the viable region to  $v_{\chi^\prime} \lesssim 14~\text{TeV}$. 
Within this domain, the masses of the heavy scalars remain bounded by  the symmetry-breaking scale, $m \lesssim v_{\chi^\prime}$, as expected  from their parametric dependence on $v_{\chi^\prime}$.

\par The analysis of the scalar spectrum consistent with the $T$ bound is  completed by including the results for the pseudoscalar mass.  In Fig.~5, we present the corresponding scan for $m_A$, which indicates  that the allowed region is restricted to  $m_A \lesssim 400~\text{GeV}$.

\par For completeness, we also evaluated the contributions of the scalar 
sector to the oblique parameters $S$ and $U$. While the corrections to 
$S$ are found to be significantly weaker than those derived from the 
$T$ parameter,  the contributions to $U$ are negligible.
\par  In Appendix A we present, in Figs.~(6) 
and (7), the results for the $S$ parameter as a function of 
$(f, v_{\chi^{\prime}}, \ldots, \lambda_9)$ after imposing the 
experimental constraint~\cite{PDG}
\begin{align}
S = -0.04 \pm 0.10 .
\label{Slim}
\end{align}

\par Similarly to the analysis performed for $T$, the left panels of these 
figures display the region compatible with the $S$ bound (light-blue 
area), while the right panels show the corresponding scalar mass 
spectrum. The upper boundary of the white distribution corresponds to 
the region where the $S$ parameter approaches its experimental upper 
limit, indicating that only very mild restrictions arise from this 
observable.

\par This behavior can be understood from the fact that $S$ is mainly 
sensitive to logarithmic mass hierarchies, whereas $T$ is directly 
affected by weak isospin breaking induced by scalar mass splittings.

\section{Conclusion}
In this work, we have carried out a systematic analysis of the scalar sector contributions of the 331RHN to the oblique parameters $S$, $T$ and $U$. While previous studies demonstrated that the gauge boson contributions become negligible at the TeV scale, our results show that the scalar sector may be strongly constrained by the oblique  parameter $T$. Our findings can be summarized as follows:

The parameter $T$ imposes strong constraints on the masses and energy scales associated with the scalar spectrum of the model. Numerical analysis indicates that for $v_{\chi^{\prime}}  \lesssim 14$ TeV
, the parameter $T$
 requires very small values of the trilinear coupling $f$, typically 
$f  \lesssim 10$ GeV.

Consequently, the scalar spectrum of the 331RHN remains close to the electroweak scale, with additional scalar states bounded by approximately 
$m_H  \lesssim 770 $ GeV , $m_A \lesssim 400$ GeV, and $m_{h_1^+}  \lesssim 800$GeV.

This structure implies that the scalar sector of the model is testable at current colliders, such as the LHC, providing a concrete opportunity to probe physics beyond the Standard Model.

In conclusion, the parameter $T$  emerges as the most restrictive electroweak observable for the 331RHN model, establishing direct correlations between the symmetry-breaking scale, $v_{\chi^{\prime}}$, and the trilinear coupling, $f$ impacting directly  the scalar mass spectrum. These results highlight the importance of electroweak precision tests in assessing the viability of extensions to the Standard Model and point to promising experimental prospects for detecting signals of the scalar sector of the 331RHN model. It is important to note that the robustness of this result depends on the decoupling of $\Phi_3$, implying that $\Phi_3$
 may significantly influence the electroweak constraints on the 331RHN framework.

 \section*{Acknowledgments}
C.A.S.P  was supported by the CNPq research grants No. 311936/2021-0.

\section*{References}
\bibliography{references}

\newpage
\appendix 
\section{}
\par  In this Appendix  we present the results of our numerical analysis for 
$S$ as a function of $(f, v_{\chi^{\prime}}, \ldots, \lambda_9)$, together 
with the corresponding  result to the scalar mass spectrum.

\begin{figure}[h]
    \centering
\includegraphics[scale=0.3]{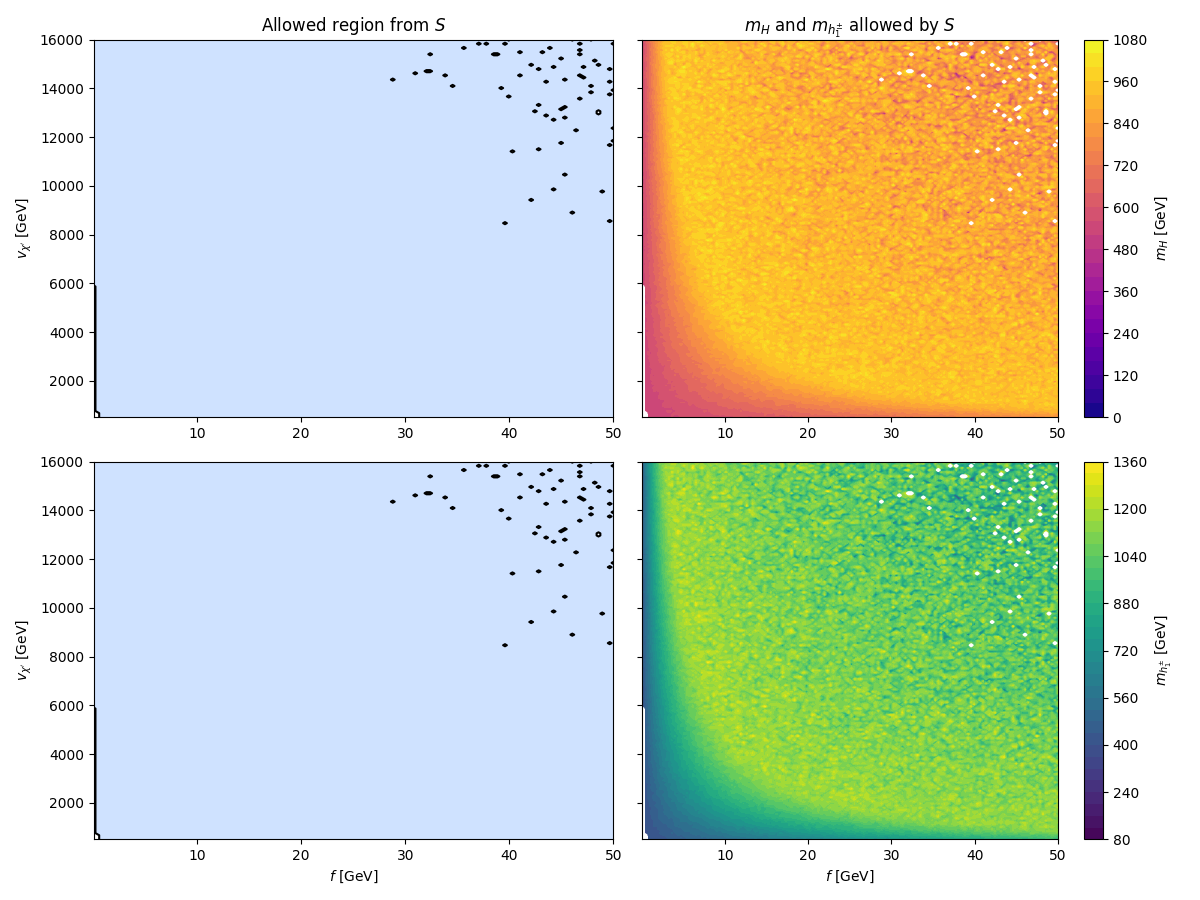}
    \caption{
Left panel: behavior of $S$ in the $(f, v_{\chi^{\prime}}, \ldots, \lambda_9)$ space after imposing the experimental constraint; right panel:  we show  $m_H$ and $m_{h_1^\pm}$ behavior, where 
the white distribution corresponds to the region where  the  $S$  parameter
reaches its experimental upper limit.}
\end{figure}

\begin{figure}[b]
    \centering
\includegraphics[scale=0.3]{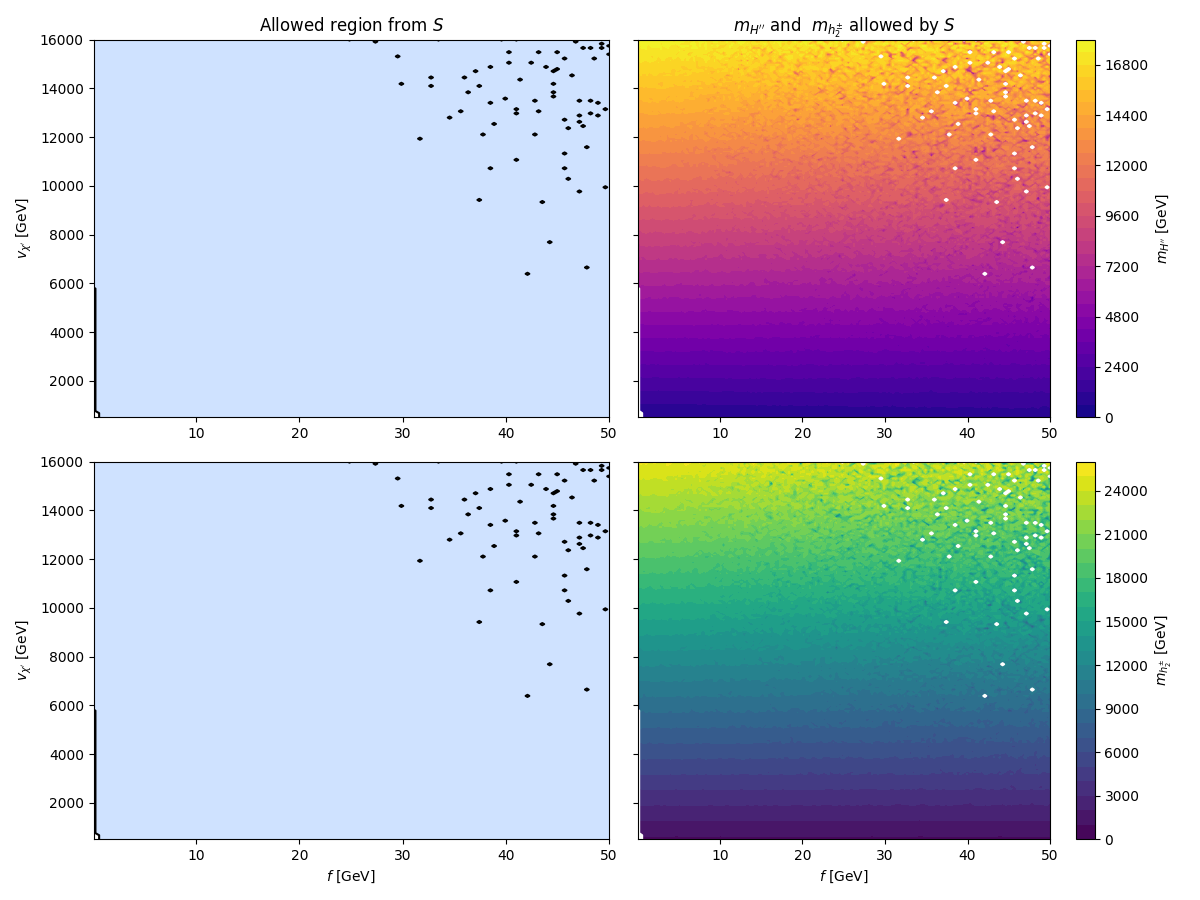}
    \caption{
Left panel: behavior of $S$ in the $(f, v_{\chi^{\prime}}, \ldots, \lambda_9)$ space after imposing the experimental constraint; right panel:  we show  $m_{H^{\prime\prime}}$ and $m_{h_2^\pm}$ behavior, where 
the white distribution corresponds to the region where  the  $S$  parameter
reaches its experimental upper limit.}
\end{figure}

\end{document}